# Enhancing Crustal Velocity Structure in Sedimentary Basin by joint inversion of Teleseismic P-Wave Reverberations and Surface Wave Dispersion

**Longlong Wang[1, 2]**

[1]State Key Laboratory of Lithospheric Evolution, Institute of Geology and Geophysics, Chinese Academy of Sciences, Beijing 100029, China

[2]College of Earth and Planetary Sciences, University of Chinese Academy of Sciences, Beijing 100049, China

## Abstract

Accurately determining the crustal velocity structure within sedimentary basins is crucial for enhancing energy resource evaluation and seismic hazard assessment. Traditional crustal imaging is challenging due to the interference of teleseismic P-wave reverberations (TPR). To address this issue, we propose an inversion strategy that combines multi-frequency TPR and surface wave dispersion (SWD) to constrain the crustal structure. Both theoretical simulations and real data tests from two stations in the Songliao Basin validate our approach. This method shows great promise for improving crustal structure investigations in complex environments, such as sedimentary basins and oceanic regions.

## 1 Introduction

Sedimentary basins are critical for understanding Earth's resource potential,

geological history, and seismic hazards. Accurately modeling their velocity structure is fundamental to deciphering basin formation mechanisms, identifying resource deposits, and mitigating seismic risks.

Traditional 2D/3D active seismic surveys provide high-resolution sedimentary structure images for petroleum exploration but are often cost-prohibitive for basin-wide coverage. Recent advances in large-scale, dense passive source arrays offer a promising alternative for obtaining 3D basin-wide sedimentary models with moderate resolution by utilizing both body wave and surface wave data.

P-to-S phases and their multiples related to the Moho (including Pms, PPmS, and PSmS) are widely used in receiver functions to image crustal thickness. However, the presence of a low-velocity sedimentary layer significantly complicates the successful application of the RF technique (Zelt & Ellis, 1999). A variety of techniques have been devised to mitigate the impact of sedimentary layers on waveforms. Once the architecture of the sedimentary layers is known, model-driven wavefield decomposition techniques (e.g., Bostock & Trehu, 2012; Chai et al., 2017; Langston, 2011; Tao et al., 2014) can effectively remove the response of the sedimentary layers from the receiver function. Additionally, some studies have developed signal processing methods to construct dereverberation filters that suppress reverberations. These methods indicate the reflection coefficient at the bottom interface, which traps the waves, along with the two-way travel time of the wave in the reverberant layer, are the key factors contributing to the emergence of reverberation signals, which can be obtained by analyzing the signal with time-domain autocorrelation (Cunningham & Lekic, 2019;

Yu et al., 2015) or frequency-domain analysis (Zhang & Olugboji, 2023)

In addition to suppressing the reverberation, waveform fitting methods have also been developed to obtain velocity structure (Anandakrishnan & Winberry, 2004; Clitheroe et al., 2000; Mandal, 2006; Sheehan et al., 1995; Zelt & Ellis, 1999), but the strong near-surface reverberations on the RFs make it difficult to reliably determine the best fitting synthetics. Despite these challenges, some research has found that multi-frequency RF inversion of the direct P phase within P-to-S converted waves (P-RFs) can provide enhanced resolution in delineating the shallow crustal structure (Wang et al., 2022).

In recent years, surface wave dispersion (SWD) has emerged as a powerful technique for investigating the absolute shear wave velocity. The development of ambient noise tomography (ANT) has further expanded the application the application of SWD, widely used for imaging multiscale S-wave velocity structures in the crust and upper mantle (Bensen et al., 2008; J et al., 2006; Li et al., 2021; Lin et al., 2008; Sabra et al., 2005; Yang et al., 2008). Joint inversion of RF and SWD can exploit the complementary information from both datasets to reduce the non-uniqueness of the solution, and has wide application in geophysics for enhanced imaging of Earth's subsurface structure  (Bodin et al., 2012; J. Julià et al., 2000; Jordi Julià et al., 2003; Shen et al., 2013; Zhao et al., 2020). However, when RF is severely influenced by the sedimentary layer, fitting both the RF waveform and the surface wave dispersion simultaneously to obtain reliable sedimentary basin structure remains a challenge.

Unscented Kalman Inversion (UKI) is a Bayesian inversion method that does not explicit derivative calculations and iteratively approximates the posterior distribution using a Gaussian distribution (Huang et al., 2022a; Huang et al., 2022b). These features make UKI computationally efficient and well-suited for the joint inversion of RF and SWD data (L. Wang et al., 2022).

In this paper, we explore the feasibility of using joint inversion of RFs and SWD with the UKI method to simultaneously determine the depths of the sedimentary basin and the Moho.

## 2 Data and Methodology

### 2.1 De-reverberation filter

The de-reverberation filter is an established method for removing reverberations from receiver functions, particularly those originating from sedimentary layers (Yu et al., 2015). This method is based on the assumption that sedimentary layer reverberations can be modeled as a diminishing sine wave pattern. Since these reverberations have a resonant frequency associated with the two-way travel time of S waves within the sedimentary layer, they can be parameterized with two physical quantities: the two-way travel time $\Delta t$ and the intensity coefficient $r_0$, which is proportional to the reflection coefficient at the bottom of the sedimentary layer. These parameters can be determined from the time-domain autocorrelation analysis of the receiver function. These analysis reveals a major negative peak with a magnitude of $r_0$ at delay time $\Delta t$.

In the frequency domain, the original receiver function ($F(\omega)$) can be viewed as the product of the signal without reverberation ($H(\omega)$) and the reverberation

$$F(\omega) = H(\omega)(1 + r_0 e^{-i\omega\Delta t}). \qquad (1)$$

By applying division in the frequency domain, the original receiver function without reverberation effect can be obtained.

## 2.2 RFs, Der RFs and SWD sensitivities to sedimentary layers and Moho depth

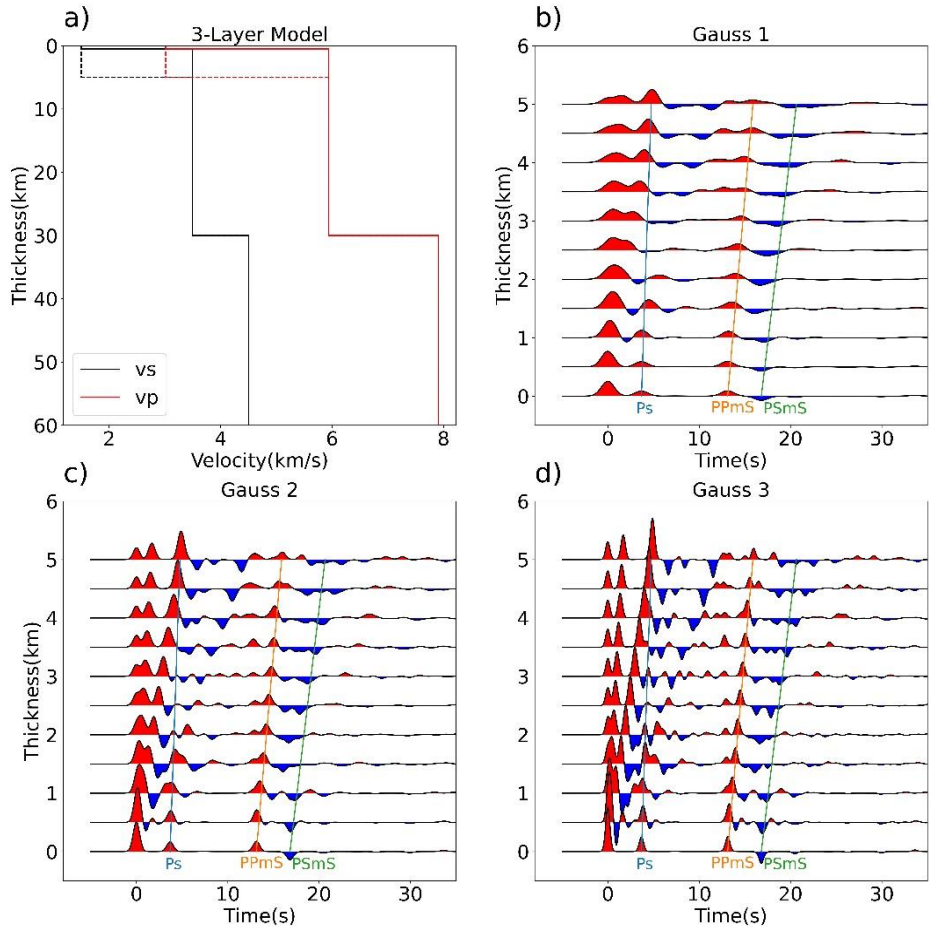

Figure 1. Synthetic examples of the effects of sedimentary layer(s) on RFs. Panel (a) shows a schematic diagram of a three-layer model with sedimentary layer thicknesses ranging from 0 km to 5 km, with the Moho fixed at 30 km. The Vp and density follow Brocher's empirical formula (Brocher, 2005). Panels (b), (c), and (d) show RFs with Gauss factors of 1, 2, and 3, respectively. The theoretical arrival times of the Ps, PPmS,

and PSmS phases are calculated using theoretical formulas (Yu et al., 2015) and are represented by blue, orange, and green lines, respectively. It can be seen that the direct Ps phase from the Moho is significantly affected by reverberations from the sedimentary layer in the RF, while the reverberation phases from the Moho, such as PPmS, are less affected by the sedimentary layer's reverberations. Therefore, the PPmS phase can be employed to constrain the Moho depth.

RFs and SWD are effective tools in seismology for probing the Earth's subsurface structures. RFs are obtained by deconvolving the vertical component from the radial component of three-component seismic records from teleseismic earthquakes. This process isolates signals related to the incoming P-phase, converted P-to-S phases (e.g., PmS from the Moho), and reverberation phases (e.g., PPmS and PSmS from the Moho), providing valuable information about sedimentary layers and the Moho depth. When a low-velocity sedimentary structure is present, RFs may be affected by reverberations generated at the base of the sedimentary layer. After applying a de-reverberation filter, these reverberations can be effectively suppressed, highlighting the reverberation phases from the Moho. SWD data (Rayleigh wave phase velocity, around 5-40s in this study) provide insights into subsurface layer properties by analyzing wave velocity variations with frequency. Numerous studies have confirmed that RFs (with Gauss factor of 1,2,3) and SWD are sensitive to sedimentary layers and the Moho depth (Gao et al., 2019; Pasyanos, 2005). In this study, we first discuss their sensitivities.

First, we set up a simple model with a single sedimentary layer and a two-layer crust.

By varying the thickness of the sedimentary layer, we can examine its effect on RFs. As shown in Fig. 1, the presence of a sedimentary layer can obscure the PS phase to some extent, but the Moho reverberation phases, particularly PPmS, are less affected by the reverberations of the sedimentary layer.

Next, we set the sedimentary layer model to a linear model and the crust to a three-layer model (with velocities of 1.5,3.5,4.5 km/s). We vary the sedimentary layer thickness to 1.5, 2.5 and 3.5 km to observe changes in the waveform of the RFs and SWD. As shown in Fig. 2 (a)-(d), the depth of the sedimentary layer affects the entire waveform of the RFs, particularly the arrival time of the PPmS phase, which becomes more evident in the de-reverberation filtered RFs (Fig. 2e).

To explore the depth of Moho to the RFs and SWD, we fix the thickness of the three layers and vary the Moho depth. As shown in Fig. 2 (f)-(j), changes in Moho depth mainly affect the arrival time of PPmS phase, additionally, these changes also impact the surface wave velocity at 10-35s, particularly around the 20s period.

We find that when a sedimentary layer is present, multi-frequency RFs with a long-time window (~20s) and SWD with 5-30s are sensitive to the depth of the sedimentary base. Additionally, RFs with periods greater than 10 seconds and SWD are also sensitivity to the Moho depth (Fig 2 (f)-(j)), particularly the PmS and PPmS phases (Fig. 2e, j). As a result, RFs and SWD (5-40s) are both sensitivitive to the sedimentary layer and Moho. Concentrating solely on the sedimentary base interface or the Moho within this dataset could result in a biased estimation of the other.

Additionally, observing Fig. 2e and f, we can see that the de-reverberation filter can suppress effect on the reverberations. Therefore, this paper explores using a joint inversion approach to simultaneously determine the depth of sedimentary base and the Moho.

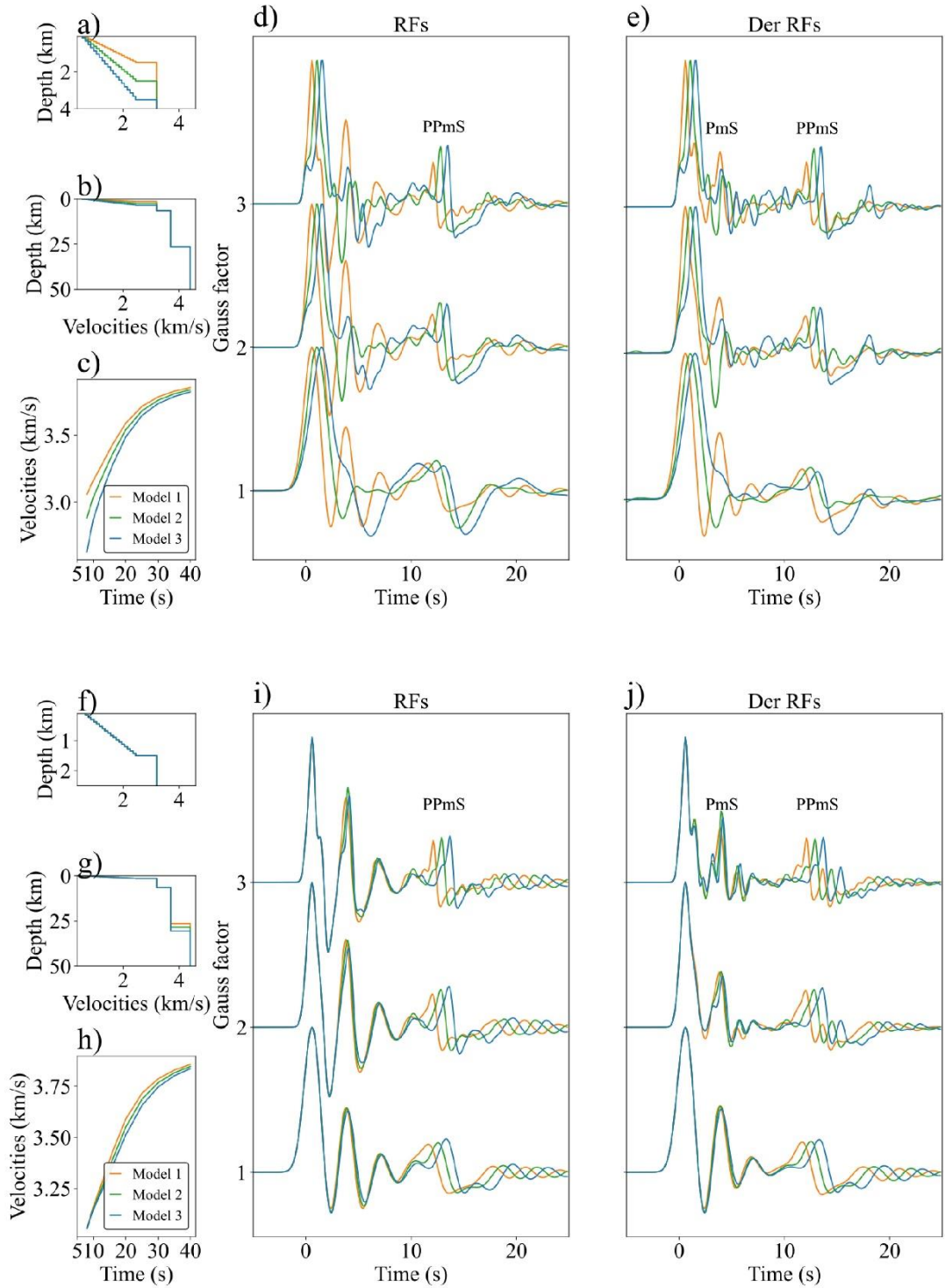

Figure 2. Synthetic examples of the effects of depth of sedimentary base and Moho on SWD and RFs. Panels (a) and (b) show the shallow and deep S-wave velocity profiles of three sedimentary layer models, with P-wave velocity and density obtained based on Brocher's empirical formulas (Brocher, 2005). Panel (c) displays the SWD data for the three models, while panel (d) shows the RFs with three different Gauss factors obtained from these models, panel (e) shows the dereverberation filtered RFs with three Gauss factors. (f)-(j) are the same for (a)-(e) but with three models that fixed the sediment and vary the different Moho depths.

### 2.3 Joint Inversion of RF, Der RF and SWD

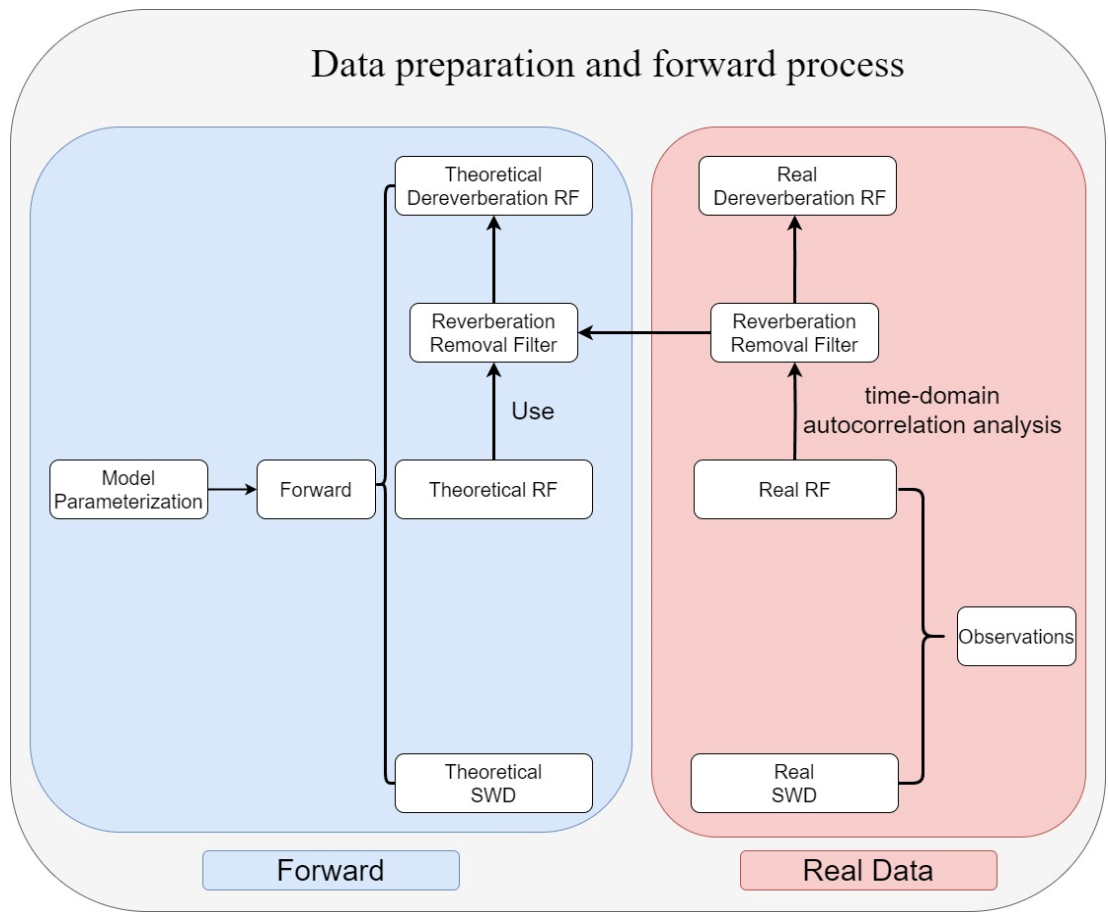

Figure 3. Data preparation and forward process within joint inversion. The right panel showcases observational data, including real SWD, real RF, and de-reverberation RF

using the de-reverberation filter from time-domain autocorrelation analysis. The left panel depicts forward modeling, involving the model parameterization, and generation of theoretical data. For simplification, only single-frequency receiver function is illustrated. In practice, multiple frequencies receiver functions are used for joint inversion.

To capitalize on the complementary sensitivities of RF data, dereverberation filtered RF (Der RF) data, and SWD, we simultaneously invert all three datasets to provide a more comprehensive constraint on basin structures. An important aspect to note is that the forward process employs the same dereverberation filter generated from the real data. This approach is adopted to generate the theoretical Der RF, ensuring consistency between the theoretical and real data in the joint inversion analysis (Fig 3).

Here we employ the UKI to perform the joint inversion (Huang et al., 2022). Unlike conventional linearization methods, UKI is a Bayesian method that iteratively updating a Gaussian distribution to approximates the posterior distribution. It strategically selects a small set of chosen sample points (sigma points (Julier & Uhlmann, 1997)) to capture the mean and covariance of the distribution. By carefully propagating these sigma points through the nonlinear system, UKI approximates a Gaussian representation of the posterior distribution, achieving greater accuracy. This method has been successfully applied to the joint inversion of RF and SWD, as detailed by (L. Wang et al., 2022). In this study, the objective function is defined as

$$p(\mathbf{m}|\mathbf{d}) \propto \exp\bigl(-\Phi_{RF}(\mathbf{m}) - \Phi_{DerRF}(\mathbf{m}) - \Phi_{SWD}(\mathbf{m})\bigr), \qquad (10)$$

where

$$\Phi_{RF}(\mathbf{m}) = \frac{1}{2}||\Sigma_{\eta_{RF}}^{-\frac{1}{2}}(\mathbf{d}_{RF} - G_{RF}(\mathbf{m}))||^2. \tag{11}$$

$$\Phi_{DerRF}(\mathbf{m}) = \frac{1}{2}||\Sigma_{\eta_{DerRF}}^{-\frac{1}{2}}(\mathbf{d}_{DerRF} - G_{DerRF}(\mathbf{m}))||^2. \tag{12}$$

$$\Phi_{SWD}(\mathbf{m}) = \frac{1}{2}||\Sigma_{\eta_{SWD}}^{-\frac{1}{2}}(\mathbf{d}_{SWD} - G_{SWD}(\mathbf{m}))||^2. \tag{12}$$

Here $G_{RF}(\mathbf{m})$, $G_{DerRF}(\mathbf{m})$, and $G_{SWD}(\mathbf{m})$ are the computed RF, Der RF and phase velocity, respectively. Meanwhile, $\mathbf{d}_{RF}$, $\mathbf{d}_{DerRF}$ and $\mathbf{d}_{swd}$ represent the observed datasets. $\Sigma_{\eta_{RF}}$, $\Sigma_{\eta_{DerRF}}$, $\Sigma_{\eta_{SWD}}$ are error covariances for each dataset. They are all diagonal matrices, with individual elements determining the relative importance of each dataset during the inversion. They are used not only for calculating the optimization error but also for controlling the weights of each dataset. In this study, considering the strong non-uniqueness of the reverberation RF inversion, we set the weight associated with SWD to be very small, at 0.001. Der RF is generated from each RF via autocorrelation analysis. Therefore, the weights of Der RF and RF are set to be the same, reflecting their similarity, and their corresponding weights are both set to 0.01.

## 2.4 Multi-prior setting

UKI minimizes the objective function eq. 10 by iteratively updating a Gaussian distribution. However, the inversion of reverberations exhibits strong non-uniqueness, a solution derived from a single prior (initial Gaussian distribution) is likely not the global optimum. To obtain more comprehensive information about the distribution of solutions, we introduce multi-prior Gaussian distributions, these prior distributions are spread as widely as possible within the potential model space and perform inversions separately. Finally, we collect the means of all resulting Gaussian distributions as the final set of solutions (The iterative result of UKI is also a Gaussian distribution, we only consider the mean value).

2.5 Forward process

We employ computer programs in seismology (Herrmann, 2013) to calculate the synthetic phase velocity and the Thomson‐Haskell propagator matrix method (Haskell, 1962; Thomson, 1950) to compute the synthetic and frequency deconvolution to calculate the RF. Der RF is obtained by applying a dereverberation filter to the RF data. We have parameterized the 1D model into two primary sections: the sedimentary layer and the crystalline crust. The sedimentary layer, defined as a Bilinear layer, is characterized by five variables: the velocity at the beginning, the velocity at the end, the transition velocity, the overall thickness of the sedimentary layer, and the proportion of the initial part relative to the entire sedimentary layer (Fig S1). The crystalline crust segment of the model encompasses two parameters: layer velocity and layer thickness. For both the sedimentary layer and the crystalline crust, the Vp and density is computed using Brocher's empirical formula (Brocher, 2005).

**3. Applications of the Method**

3.1. Application to synthetic data

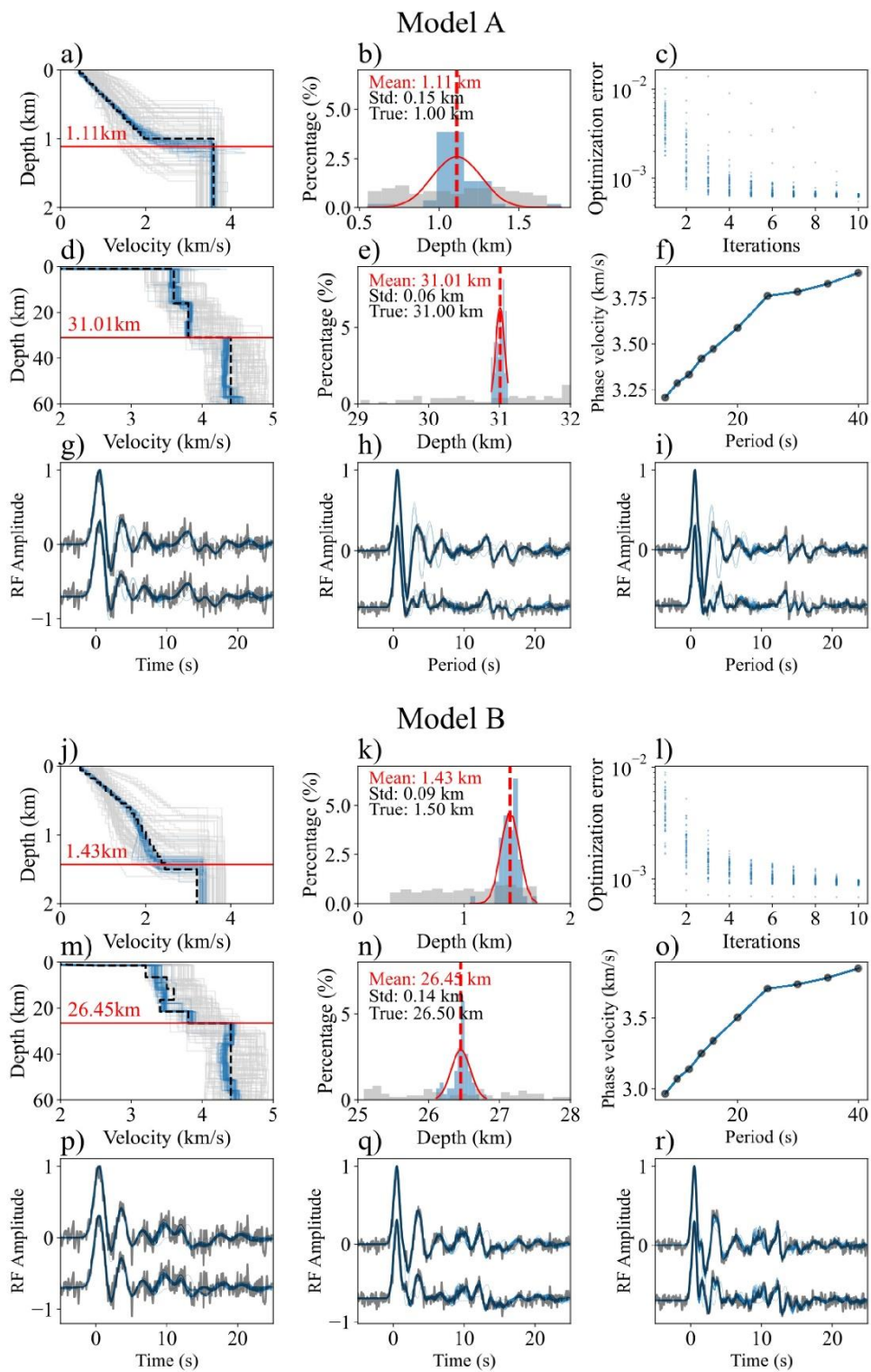

Figure 4. Inversion analysis of theoretical sedimentary Model A, B. (a-i) Results for the

Model A: (a, d) Velocity structures with the theoretical model in black. The 200 initial

models (mean of the initial distribution) are shown in gray lines, while Top 30%

inversion results with the lowest optimization errors are shown in light blue. Red lines highlight the good results' mean sedimentary layer depth and Moho depth. (b, e) Depth statistics for the sedimentary layer and Moho. (c) Iteration errors. (f) Phase velocity comparison between inversion results and observations. (g-i) Waveform fits for Gaussian coefficients 1-3. Original waveforms are centered around the vertical coordinate 0, while post-reverberation waveforms are centered around -0.7. (j-r) Similar results as above, but for the Model B.

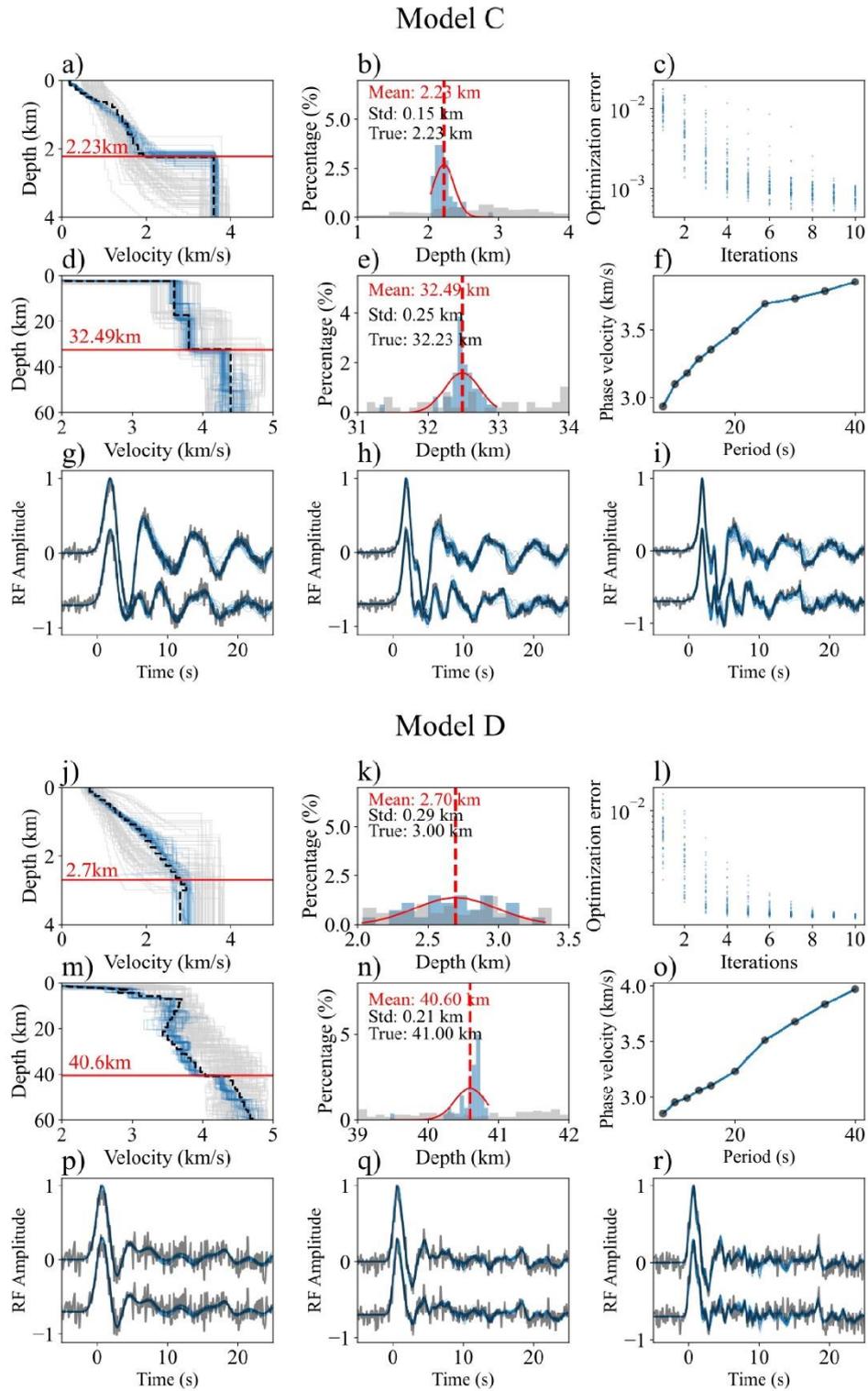

Figure 5. Same as Fig. 4 but for Model C and D. Model C features a sedimentary structure that is not a bilinear structure, obtained from the results of real higher-mode SWD (Xie et al., 2023), while Model D has a mid-crustal gradient low-velocity layer.

To illustrate the feasibility of our proposed joint inversion, we offer a detailed inversion result of 4 theoretical models. Model A features a simple linear sedimentary model combined with a three-layer crustal model. Model B has a bilinear sedimentary model with a low-velocity layer in the crust. Model C's sedimentary obtained from the inversion results of real higher-mode SWD (Xie et al., 2023). Model D includes a gradient low-velocity structure in the crust. All models generate theoretical RFs with Gaussian factors of 1, 2, and 3 through the forward function (Eq. 2.4), adding Gaussian white noise with a standard deviation of 0.04. We then apply time-domain autocorrelation analysis to produce the corresponding de-reverberation filters, obtaining de-reverberated RFs, and incorporating the filters into the forward function (Fig. 3).

Before performing the inversion, we need to consider the parameterization of the model. We parameterize the sedimentary layer with five parameters (Fig. S1) and utilize the reference model (the gradient-segmented AK135 model) as the basis for the crystalline crust. The sedimentary layer model is then combined with the crystalline crust model to form a complete initial model.

Unlike traditional gradient-based methods which update a single model with each iteration, UKI iteratively updates a Gaussian distribution. Therefore, in addition to setting the initial mean model, it also requires setting the initial covariance matrix. In this study, we set the covariance matrix uniformly as a diagonal matrix with values of 0.05 (Huang et al., 2022; L. Wang et al., 2022).

To prevent the results from falling into local minima due to inappropriate priors, we introduce random perturbations to prior Gaussian mean to get 200 prior distributions.

These distributions are assumed to broadly cover the potential model space. For each prior distribution, we perform an inversion with 10 iterations, adjusting the priors by incorporating observational data to obtain a posterior that aligns with the observed data.

Ultimately, we derive 200 posterior Gaussian distributions. By averaging their mean values, we generate a final comprehensive distribution of results.

Fig. 2 demonstrates our recover tests for the Model A, with red lines pinpointing the mean sedimentary layer thickness and Moho depth. Panels (a) and (d) display the velocity structures, contrasted against our theoretical models (in black). Detailed depth statistics for both the sedimentary layer and Moho are provided in panels (b) and (e).

Panel (c) illustrates how iterative error is reduced during the analysis. Crucially, panel

(f) demonstrates our method's accuracy by comparing inversion results with observed

phase velocity data. Panels (g-i) showcase Gaussian coefficients 1-3, highlighting waveform changes. Notice the original waveforms aligned near vertical coordinate 0, compared to the distinct skew towards -0.7 in the post-reverberation waveforms.

To verify the stability of our method, we also selected Models B, C and D, and their

corresponding noise-added theoretical observation data for inversion tests, as shown in

Fig. 3 and 4. These results show that our method can also effectively recover the sedimentary layer and Moho depth for these models.

### 3.2. Application to actual data

We apply our inversion method to analyze data from two broadband seismic stations within the Songliao Basin of northeast China. The first station, NE68, is situated within the Changling graben-depression (Changling sag), where significant sedimentary accumulation reaches up to 9 km. With approximately 510 meters of unconsolidated Quaternary and Neogene sediments (Feng et al. 2010), NE68 operated from September 2009 to August 2011, and provided waveform data from 305 teleseismic events. The second station, NE96, is located in the basin's northern region near the Daqing (Taching) Oil Field. Both stations were part of the NECESSArray project (Steve & Jim, 2009), which ran from September 2009 to September 2011.

We obtain the phase velocities in the periods of 8‑40 s of the two stations from Guo et al. (2015), respectively. We select teleseismic records at the two stations from earthquakes with a magnitude greater than 6.0 and an epicentral distance between 30° and 90°. Radial RFs are calculated using time-domain iterative deconvolution with specific Gaussian parameters ($\alpha = 1, 2, 3$), and we use $\alpha = 3$ as the reference to select the RFs based on rules in Supplementary S4, resulting in the Selected RF data. The data was then categorized by ray parameter and presented as a bin chart, as shown in Fig. 6. This illustration provides a detailed depiction of the RF with $\alpha = 3$ and associated waveforms for two stations. Three ray parameter blocks, distinguished by blue, purple, and green, will be chosen to calculate the mean RF separately. Then RF with $\alpha = 1,2$ will also be calculated at the same ray parameter block.

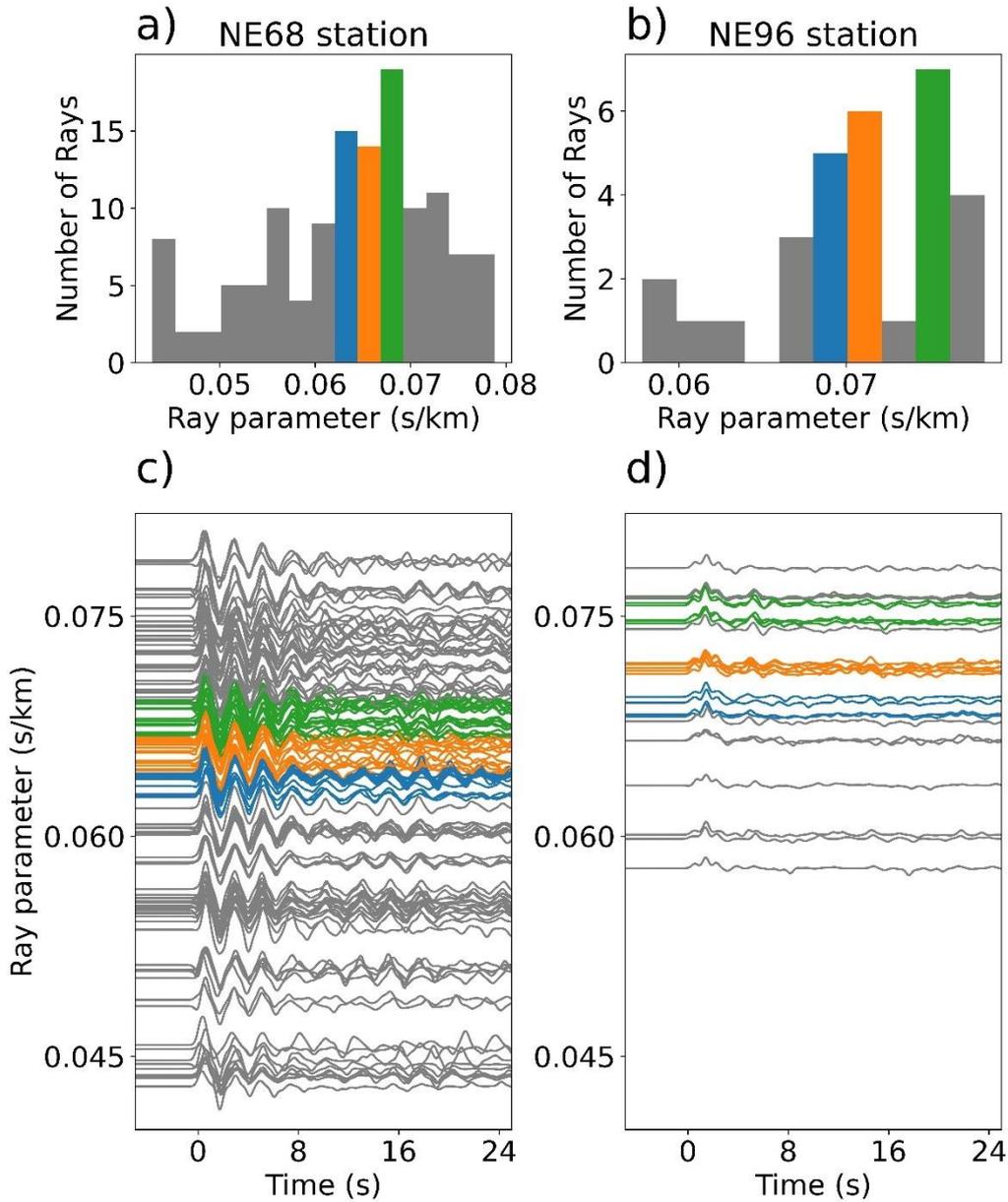

Figure 6. Selected Receiver Functions (Only show the RF with Gaussian coefficient of

3) and Waveforms for Stations NE68 and NE96. (a) Distribution of receiver functions

at NE68 by ray parameters, with predominant bands marked in blue, purple, and green

for subsequent mean RF calculations. (c) NE68 waveforms categorized by ray

parameters. Key bands are emphasized in corresponding colors for focused mean RF

analysis. (b)(d) Same as (a) and (c), but for NE96 stations.

　Follow the UKI setting above, we set the 200 initial Gaussian distributions

(differing only in mean values, plotted in grey in Fig S2), then we obtain the 60 good results (Top 30% according to optimization error) for each ray parameter, this approach was adopted to demonstrate the feasibility and stability of our method. To provide a clearer representation of the outcomes, we detail the corresponding inversion results in Fig. 7 and 8.

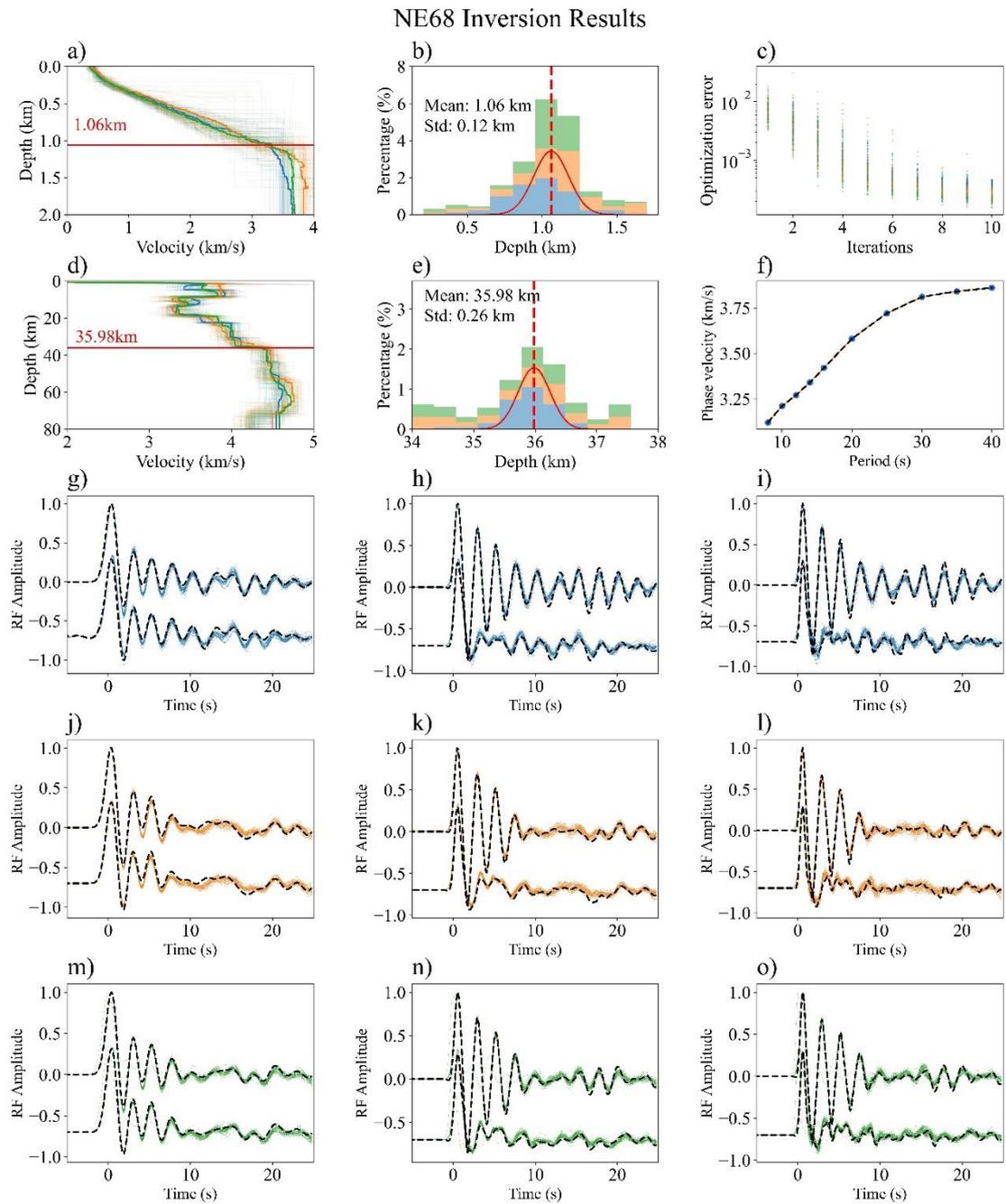

Figure 7. Inversion results for RFs and SWD at station NE68 with 3 ray parameters. (a,

d): Sedimentary and crustal structures color-coded by ray parameters. Lighter shades indicate 30% of inversion results with the lowest optimization errors, the darker shades represent the average model for each color, while faint gray lines represent the initial models. Red lines delineate the good models' mean sediment depths and the Moho depth. (b, e): Depth statistics for the sedimentary layer and Moho. (c): Optimization error for good models. (f): Comparison with actual SWD data. (g-i, j-l, m-o): Inversion results for three specific ray parameters, showcasing the method's capability in revealing subsurface details amid sediment effects.

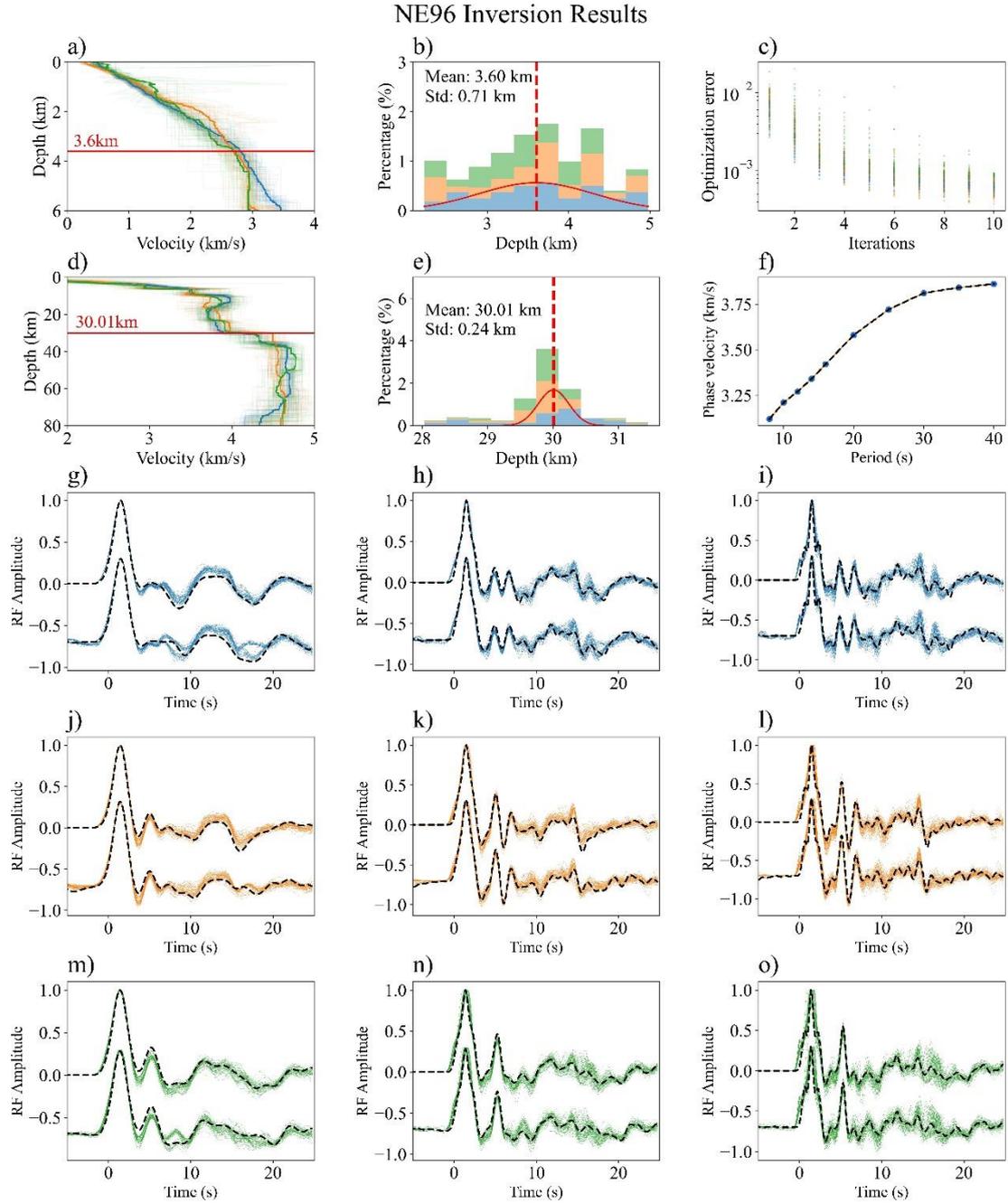

Figure 8. Inversion results for receiver functions (RF) and surface wave dispersion (SWD) at station NE96 with 3 ray parameters. It can be observed that the de-reverberation filter does not significantly affect the waveforms of real data.

The inversion results for RFs and SWD at station NE68 are displayed in Fig. 7, each corresponding to one of the three distinct ray parameters in Fig. 6. The sedimentary and crustal structures are portrayed in panels (a) and (d), where colors are mapped to ray parameters, and initial models are denoted in faint gray. Depth profiles, derived from the top-performing models, are elucidated in panels (b) and (e). The optimization error is showed in panel (c), while panel (f) emphasizes our alignment with the observed surface wave dispersion data. Despite minor waveform variations in RFs across ray parameters, their inversion results demonstrate substantial alignment—particularly in delineating sedimentary layer thickness and Moho depth, as shown in panels (b) and (e), which further corroborates the stability of the method. Moho depth of NE68 is about 35.85 km, which is consistent with previous study by Yu et al., (2015) with 35 km. We define the bottom boundary of the bilinear model as the interface between the sedimentary layer and the bedrock, which is about 1.06±0.12 km (as seen in Fig. 7). The S wave velocity of the sediment layer starts at approximately 0.4 km/s, indicating predominantly unconsolidated sediments at this site. Additionally, our results reveal a very low-velocity layer in the shallow part, which aligns to some extent with the 0.35 km thickness of the sedimentary layer reported by Yu et al., (2015). Additionally, within our inversion results, we identified a low-velocity layer situated between 10-20 km depth, exhibiting a velocity of approximately 3.2 km/s. This low-velocity layer consistently emerged across inversions from three ray parameters, each producing good waveform fits. This consistency strongly suggests the existence of the low-velocity layer. Zhan et al., (2020) employed multimodal inversion of Rayleigh waves from ambient seismic noise and similarly identified a significant low-velocity layer in the mid-crust beneath the Songliao basin.

Fig. 8 presents the results from the NE96 station. One of the key features at this station is the noticeable delay in the arrival of the direct P-wave. The inversion results shown in Fig.8, with various ray parameters represented by distinct colors, provide a comprehensive overview of the subsurface structures. NE96's sedimentary layer has an approximate depth of ~3.6±0.71 km, and an Moho depth about 30.01 km. Additionally, studies using frequency dependent P wave particle motion have determined that the sedimentary layer thickness at station NE96 is approximately 4.2 km(Bao & Niu, 2017), which is consistent with our findings. When we convert the S-wave velocity to P-wave velocity, the calculated P-wave double travel time is approximately 10.5 seconds, aligning with the results for the Moho from wide-angle reflections (Yang, Baojun, 2003).

## 4. Discussion

In this study, we have designed a strategy for inverting teleseismic P-wave reverberations, that joint inversion with SWD. This section will focus on discussing the setting of initial parameters, validating the accuracy of our results, examining the soundness of our methodological assumptions.

To prevent the UKI inversion from falling into local minima, we set multiple initial models and invert them separately as described in section 2.4. This approach ensures a thorough exploration of the solution space and increases the likelihood of finding a robust solution. To enhance the efficiency of the inversion process, we can use waveform information to provide more accurate priors, such as the delay time of P-wave arrivals (Bao & Niu, 2017; Deng et al., 2023; Marignier et al., 2024). Another factor that requires consideration before inversion is the number of model parameters.

When the data quality is poor, we recommend reducing the number of model parameters and increasing the interval of layer thickness. Conversely, when the signal quality is good, we suggest increasing the number of model parameters and decreasing the layer thickness interval. For example, the data from Model C is heavily affected by reverberation, so we only use 2 layers to represent the crystalline crust. In contrast, the data generated by Model D is less affected by reverberation, so we add the model layers in the crust and use 4 layers to represent. According to Occam's razor principle, if the primary focus is on the depths of the sedimentary layer and the Moho, it is recommended to fit the waveform and dispersion with as few model parameters as possible.

To demonstrate that the results obtained from real data are reliable, we begin with a model recoverability test of the joint inversion. Using an approximation of the inversion results of the NE68 receiver functions as the input model, we obtain the observation datasets from the forward process. Keeping the inversion setting consistent, we perturb the initial mean 200 times to inversions. The top 30% of the results are selected. For enhanced visualization, we use a Gaussian distribution to approximate the distribution of results. As shown in Fig. 9a, the inversion outcome demonstrates a goot fit with the input model.

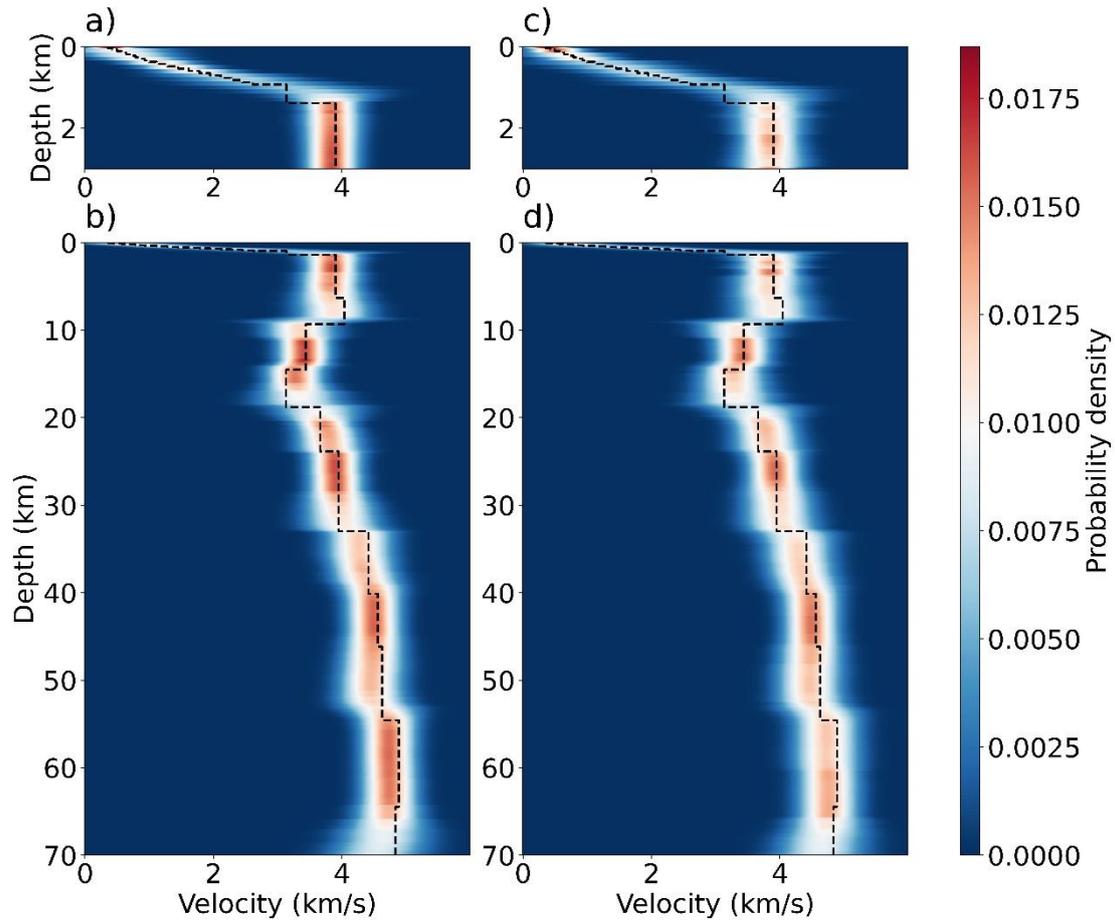

Figure 9. Compares the probability distributions of the inversion results (Top 30% of 200 results) with and without the dereverberation filter in the joint inversion. (a) and (b) show the probability distributions of the inversion results with the dereverberation filter. (c) and (d) show the probability distributions of the inversion results without the dereverberation filter.

To verify the effectiveness of the de-reverberation filter, we conducted a test with the same theoretical model, using only reverberation RF and SWD for joint inversion. We kept the inversion settings consistent and performed 200 inversions with the same perturbed initial means, selecting the top 30% (lowest optimization errors) of the results. We then compared these results with those from the previous recovery test. As shown in Fig. 9b, with the probability density exhibits a significantly higher peak value concentrated around the input model. This demonstrates that de-reverberation filter can enhance the accuracy of the inversion results. Although the de-reverberation filter may not be effective when the sedimentary layer is relatively thick (as shown in Fig. 8), we recommend including it in the inversion process for consistency. When MCMC or other methods are used to invert reverberation in the future, we also recommend incorporating a de-reverberation filter to enhance efficiency and obtain more reliable results.

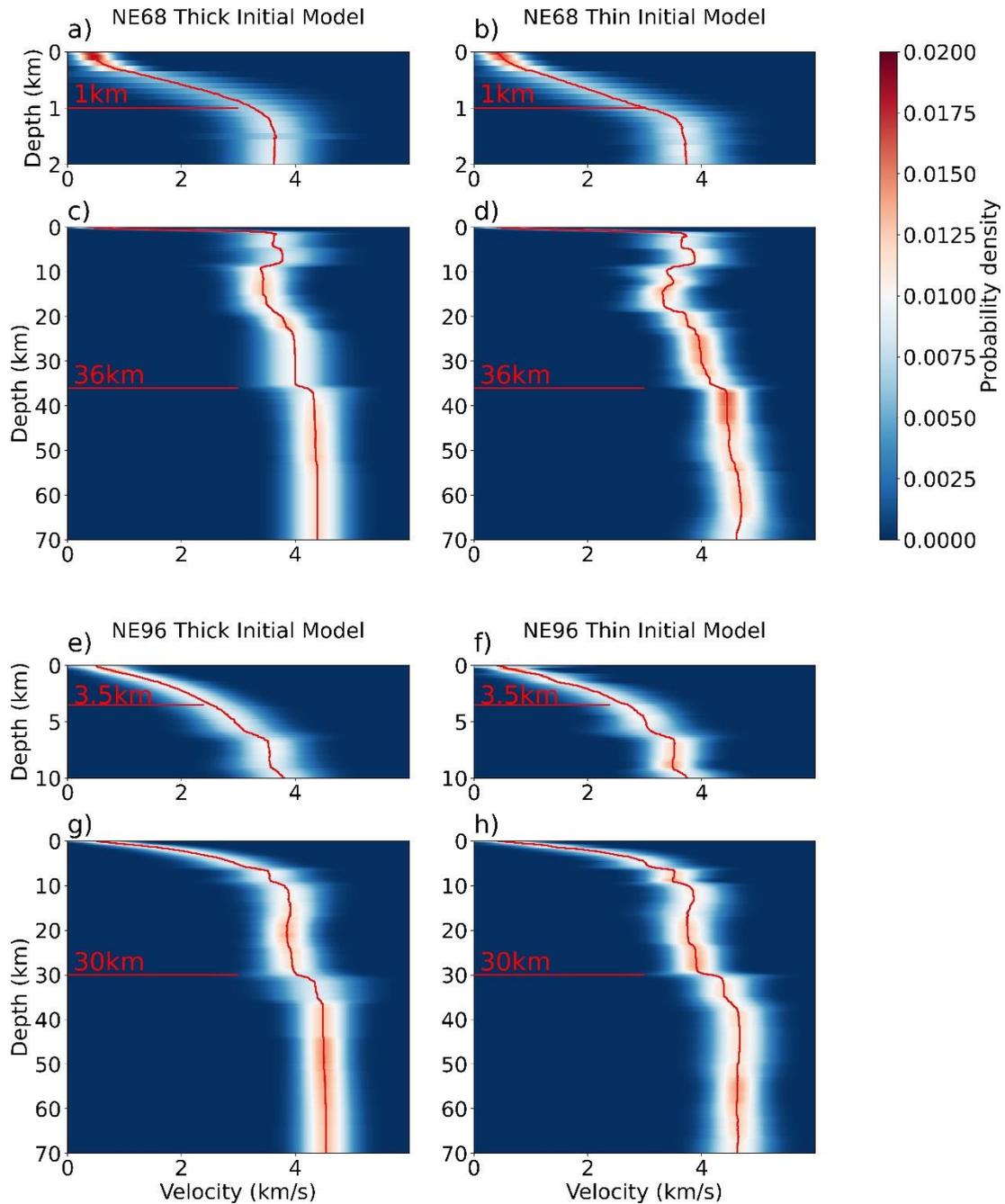

Figure 10. Comparing inversion results based on initial distributions with differing layer thicknesses. (a) and (b): Inversion result probability distributions (30% lowest optimization errors) using initial distributions with mean values from Fig. S2(a). (c) and (d): Inversion result probability distributions (30% lowest optimization errors) using initial distributions with mean values from Fig. S2(b). (e)-(h) are the same as (a)-(d) but for NE96 stations.

To investigate the influence of the initial thickness on the result, we compared the inversion results obtained using relatively thin and thick initial models in crust, as shown in Fig. 10, the inversion results of the two models are not significantly different and show a high degree of consistency in the recovery of the sedimentary structure and Moho depth.

For future work, using higher frequency receiver functions, higher-order surface waves, and employing higher-dimensional sedimentary models for inversion will be a growing trend.

## 5. Conclusion

The presence of sedimentary layers generates strong reverberations in receiver functions (RFs), making waveform fitting particularly challenging. The primary objective of this study is to demonstrate that utilizing high-quality, multi-frequency, long-window RFs (ranging from -5 to 20 seconds) and SWD data (Rayleigh wave phase velocity, around 5 to 40 seconds can effectively constrain the crustal structure. Our approach has been validated through both theoretical models and real data, underscoring the feasibility and effectiveness of this inversion strategy. Although we employed unscented Kalman inversion (UKI) to support our conclusions, we believe that other inversion methods, such as Markov Chain Monte Carlo (MCMC), are also viable and could provide reliable results.

**Data Availability Statement**

RFs data and the SWD data from the station KIGAM can be accessed through the CPS tutorial (http://www.eas.slu.edu/eqc/eqc_cps/TUTORIAL/STRUCT/index.html).

**Acknowledgements**

Longlong Wang is deeply grateful to Dr. Yun Chen for his considerate guidance and support throughout the research process. He also extends his appreciation to Dr. Youshan Liu for the substantial improvements he has made to the manuscript. Furthermore, Longlong Wang is thankful for the enlightening conversations and constructive advice provided by Dr. Zhengyu Huang. With their consent, upon the publication of this paper, they will be acknowledged as co-authors for their significant contributions.